# Systems biology approach to the origin of the tetrapod limb

Koh Onimaru[1*], Luciano Marcon[2*]
1. Laboratory for Phyloinformatics, RIKEN Center for Biosystems Dynamics Research (BDR), 2-1 Hirosawa, Wako city, Saitama prefecture, 351-0198, Japan. E-mail: koh.onimaru@riken.jp
2. Andalusian Centre for Developmental Biology (CABD), Universidad Pablo de Olavide, Carretera de Utrera km1, 41013, Sevilla, Spain. E-mail: lmarcon@upo.es
*Corresponding author

July 2019

**Abstract**
It is still not understood how similar genomic sequences have generated diverse and spectacular forms during evolution. The difficulty to bridge phenotypes and genotypes stems from the complexity of multicellular systems, where thousands of genes and cells interact with each other providing developmental non-linearity. To understand how diverse morphologies have evolved, it is essential to find ways to handle such complex systems. Here, we review the fin-to-limb transition as a case study for the evolution of multicellular systems. We first describe the historical perspective of comparative studies between fins and limbs. Second, we introduce our approach that combines mechanistic theory, computational modeling, and *in vivo* experiments to provide a mechanical explanation for the morphological difference between fish fins and tetrapod limbs. This approach helps resolve a long-standing debate about anatomical homology between the skeletal elements of fins and limbs. We will conclude by proposing that due to the counter-intuitive dynamics of gene interactions, integrative approaches that combine computer modeling, theory and experiments are essential to understand the evolution of multicellular organisms.

**Introduction**
Multi-cellular organisms have evolved a remarkable diversity of their forms. In the last decades, thanks to the advance of DNA sequencing technologies, the genome sequence data of various species have become available. However, we still poorly understand how genome sequences are connected to the diverse morphologies of animals. One of the causes that make genotype-phenotype mapping so difficult is that the dynamical behaviour of developmental systems is not fully encoded in the genome [1]. Indeed, developmental systems involve numerous gene interactions that can generate emergent phenomena which cannot be intuitively explained by knowing just the list of genes involved – or even their interaction network. An example of such phenomena is the diffusion-driven instability proposed by Turing [2], in which cross-regulatory interactions between multiple diffusive molecules spontaneously generate periodic spatial patterns. Therefore, in order to map genotypes into phenotypes, we need to understand the causal relationship between genetic mutations and the developmental phenomena that are not directly encoded in the genome.

The difficulty of genotype-phenotype mapping is also related to the homology problem. Homology is fundamental to evolutionary studies, since it provides the frame of reference to identify differences between species. Traditionally, homology has been defined according to two different notions [3]:
- "*the correspondency of a part or organ, determined by its relative position and connections, with a part or organ in a different animal*" [4],
- "*Attributes of two organisms are homologous when they are derived from an equivalent characteristic of the common ancestor*" [5].

The first one is Richard Owen's original definition, which was proposed before the development of the evolutionary theory. The second is Mayr's definition, which was supposed to replace Owen's one in the context of modern evolutionary synthesis. Nowadays however, homology has often been associated with both notions [3, 6]. As pointed out by Wagner [3], although the concept of homology was first defined to systematically categorize anatomical modules, comparative analysis of coding genes is the most successful and practical application of homology. Because genes are one-dimensional sequences of nucleotides or amino acid, they are algorithmically comparable, which allows the quantitative assessment of Owen's homology. In addition, because genes are directly inherited from ancestors via replication, comparison of genes is conceptually straight forward in terms of the continuity from the common ancestor. In contrast, anatomy is often about complex three-dimensional structures, thus it is difficult to assess homology with Owen's definition, and it is reconstructed from genome information at every generation through complex developmental dynamics, which makes it difficult to trace back homology by applying Mayr's definition. Without understanding of how genome information is translated into anatomical traits, homology could be concluded from superficial resemblances with ambiguous ancestral continuity. Therefore, a proper description of the genotype-phenotype map is required to solve the homology problem and to provide a better conceptual basis for anatomical comparisons.

In this chapter, we will use vertebrate fins and limbs as a case study for the evolution of multicellular systems, showing how understanding developmental mechanisms demystifies the problems on comparative analysis between fins and limbs. Firstly, we review a historical perspective of studies of the fin-to-limb evolution. Secondly, we discuss positional information and the Turing mechanism, two key developmental concepts involved in the skeletal pattern formation of the tetrapod limb. Thirdly, we will introduce a recent effort to understand molecular mechanisms that differentiate fins and limbs.

**Homology debates on fins and limbs**
The tetrapod limb is often divided into three anatomical modules: stylopod (humerus/femur), zeugopod (ulna and radius/fibula and tibia), and autopod (wrist and digits), ordered from proximal to distal (Fig. 1A). Most limb variations are explained by reduction or elongation of the skeletal elements of this basic architecture, which Richard Owen called "archetype". However, comparison between fish fins and tetrapod limbs has been controversial for more than 150 years. First of all, to clarify the terminologies, fish fins are generally composed of "radials" (endoskeletal elements) and "rays" (the fluttering part of fins made of dermal bones in ray-finned fishes or stiff filaments in the case of sharks). Here, we mainly discuss fin radials, because rays are thought to have been lost in the tetrapod lineage. A German zoologist, Gegenbaur was the first to seriously tackle this problem [7]. Although his main hypothesis that fins/limbs originate from gill arches is no longer supported (reviewed by Jarvik [8]), his attempt to identify homologous elements between fins and limbs is still influential in contemporary research. He categorized the adult skeletal elements of a shark pectoral fin into three modules along the anterio-posterior (AP) axis: propterygium, mesopterygium, and metapterygium, each connected to a large basal element (Fig. 1B). While fish fins generally exhibit considerable variation between species, he identified that the metapterygium was the most constant module across fish fins (see Fig. 5E for some fins). The metapterygium is characterized by its unique branching pattern—multiple small skeletal elements are connected to a series of thick ones (later named as the metapterygial axis; orange line in Fig. 1B). Because of the robustness of the mepterygium, he speculated that the tetrapod limb also originated from the metapterygium, and drew a hypothetical metapterygial axis on the skeletal elements of the tetrapod limb (Fig. 1C). Since then, many researchers have attempted to identify the metapterygial axis in tetrapod limbs by incorporating embryological and fossil data, but never reached a consensus. In particular, while the majority of researchers consider that the axis "runs" through humerus and ulna (femur and fibula for the hindlimb), there was a strong disagreement regarding the autopod, which promoted a debate whether digits are the *de novo* structure of tetrapod limbs or just modified fin elements. For example, Watson, 1913 suggested that the axis run through the digit IV, therefore digit I to III are corresponding to the preaxial fin radials and digit V is the postaxial fin radial [9]. On the other hand, Holmgren, 1952 suggested the axis stop running around the wrist, so the autopod emerged as a *de novo* structure in the tetrapod lineage [10]. These significant variation between different hypotheses raised doubt about the conceptual validity of metapterygial axis [8].

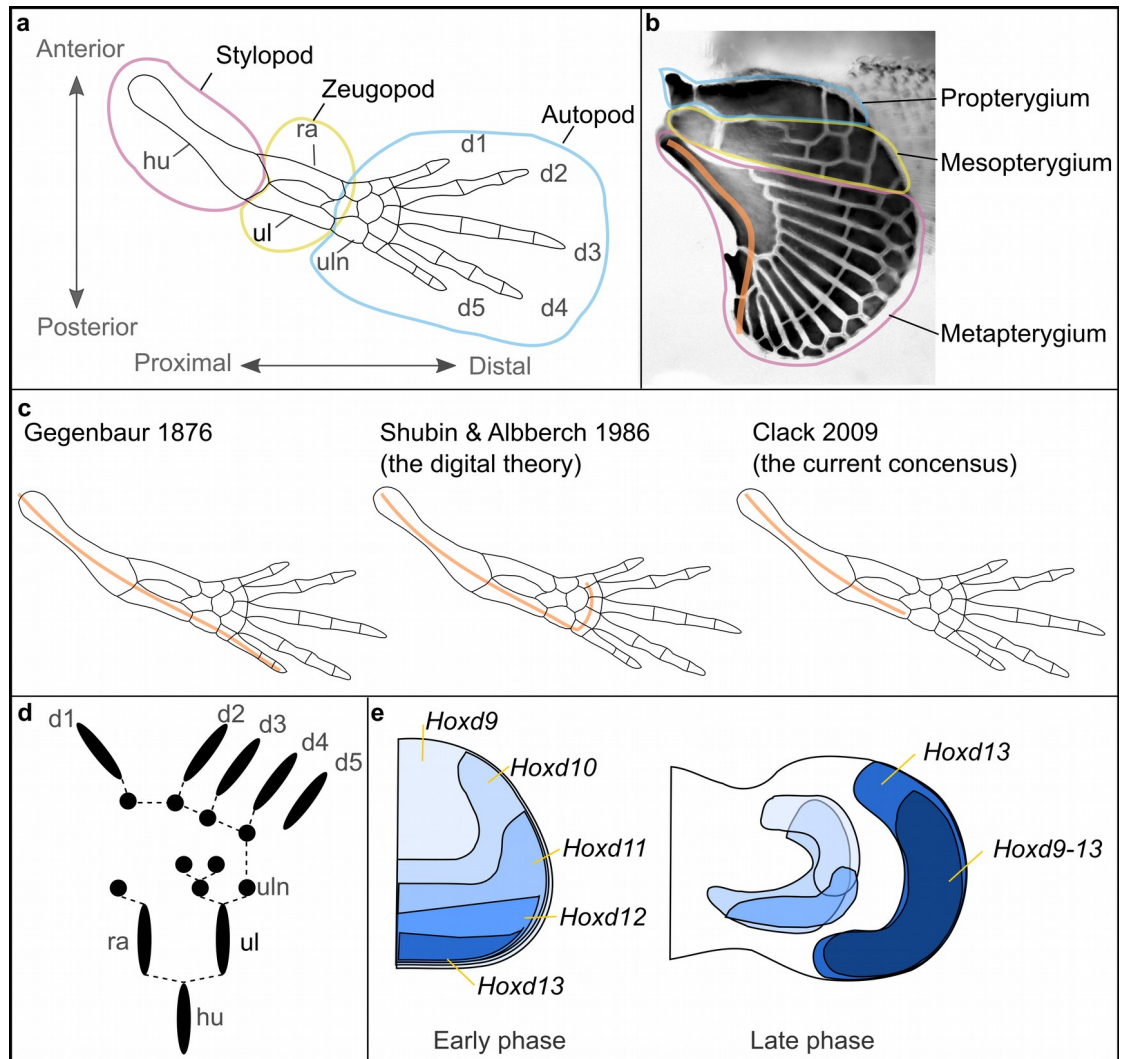

Figure 1: The history of comparative study of fins and limbs. **a**, The tetrapod limb. **b**, A pectoral fin of the small-spotted catshark. **c**, Metapterygial axes. **d**, Shubin and Alberch's scheme of a mouse forelimb, adapted from Shubin & Alberch, 1986 [6]. **e**, The expression patterns of *Hoxd* genes in mouse limb buds, adapted from Tarchini & Duboule 2006 [11]. hu, humerus; ul, ulna; ra, radial; uln, ulnare; d1-d5, digit 1-5.

Even though all of the attempts to find homologous structures between fins and limbs failed to reach a solid conclusion, one theory became highly influential: the digital arch theory proposed by Shubin and Alberch, 1986 [6]. Although this theory has been largely rejected by experimental data [12, 13], it involves a careful conceptual development of homology. Shubin and Alberch attempted to reexamine the concept of homology on the basis of developmental mechanisms by viewing limb development as "*the product of a combination of 'global' organizers ... and 'local' interactions that characterize the process of chondrogenesis*". In general, chondrogenic condensation in limb development gradually and continuously progresses from proximal to distal. When a new element starts forming, chondrogenic condensations either branches-off or segment from the edge of already existing condensations. Shubin and Alberch considered that these local processes were necessary mechanisms for skeletal pattern formation. In addition, they assumed that the type of bifurcations that occurs (i.e., branching or segmentation) was determined by a global organizer, such as the ZPA (discussed later). Their meticulous comparison of various tetrapod limbs, lungfishes and other fossil fishes (Fig. 1D as mouse example) led them to propose three conclusions: a) branching events are involved only in the postaxial elements, b) the postaxial elements also give rise to the digits via branching from the distal carpals (wrist bones), called the "digital arch", c) the postaxial elements including the digital arch, is homologous to the metapterygial axis. Therefore, according to this theory, tetrapods acquired the autopod domain by bending metapterygial axis anteriorly during evolution (Fig. 1C).

The theory subsequently received supportive data from the analysis of genes related to positional information—*Hox* genes. During mouse limb development, 5′*Hoxd* genes (*Hoxd9*, *Hoxd10*, *Hoxd11*, *Hoxd*12, and *Hoxd*13) exhibit mainly two phases of spatial expression patterns [11, 14, 15]. In the early phase, *Hoxd9* to *Hoxd13* are transcribed in a collinear manner; *Hoxd9* is expressed in the whole limb bud, and the expression domains of d10 to d13 are restricted posteriorly in a nested manner. In the late phase (during the autopod formation), the expression domain of *Hoxd13* is expanded anteriorly to cover all the digit forming regions, and those of the others only overlap with digit II to V (Fig. 1E). Because the anterior expansion of *Hoxd* gene expression seemed correlated with the hypothesis that the axis was bent to the anterior side, this late phase of *Hoxd* gene regulation was interpreted as molecular evidence for the digital arch [16]. This interpretation was also supported by an observation that zebrafish pectoral fins lacked the late phase of Hox gene expression [17]. However, subsequent examinations of several other fish fins and re-examination of zebrafish fins showed that the anterior expansion of *Hoxd* gene expression is a deeply conserved mechanism across jawed vertebrates [18–21]. As discussed later, these distally expressed *Hoxd* genes in fish fins seem to regulate the formation of fin rays (dermal bones) [22, 23] (however, this is still controversial [24]). Therefore, the late phase of *Hoxd* gene expression does not provide strong evidence for the digital arch theory.

Indeed, the theory has been often dismissed [12, 13]. The weakness of this theory is that the branching and segmentation processes during skeletal pattern formation is an observed empirical regularity rather than a real developmental mechanism. A critical observation that contrasts with the digital arch theory is that the autopod can form independently of the proximal modules without branching processes [13]. Relying on empirical regularities to identify homology is after all the same approach to Gegenbaur's analysis. However, dismissing the digital arch theory does not rule out the existence of the metapterygial axis. Indeed, current paleontological data suggest that the axis runs through at least the humerus/femur and the ulna/fibula [25] (Fig. 1C). On the other hand, other recent molecular evidence also suggests that the loss of the propterygium and the mesopterygium during the fin-to-limb transition resulted from a fusion of the three modules [26], which roughly congruent with Jarvik's argument based on the analysis of spinal nerve distribution in fins and limbs [8]. This questions the general practicality in defining homology between individual skeletal elements of fins and limbs with a traditional homology framework. Instead, the concept of homology should be focused on discovering which developmental mechanisms underlie the weak identity of the skeletal elements of fins and limbs. Therefore, in the next three sections, we review the two mechanisms that have been proposed to explain skeletal pattern formation. One is the positional information model, where patterning is controlled by a global asymmetric organizer, and the other is the Turing mechanism, where patterning results from local interactions.

**Global asymmetric organization in limb development**

Embryonic limb development is a classic model of developmental biology, serving as a conceptual platform to understand morphogenesis in multicellular systems. In particular, it is a traditional model system to explain pattern formation on the basis of positional information by morphogen gradients. In the mid 20th century, Saunders and his colleague discovered two signaling centers in chick limb development—the apical ectodermal ridge (AER; [27]) and the zone of polarizing activity (ZPA; [28]). These findings contributed to the early conceptual development of positional information, which proposed that morphogen gradients generate spatially heterogenous cell differentiation, proposed by Luis Wolpert in 1969 [29].

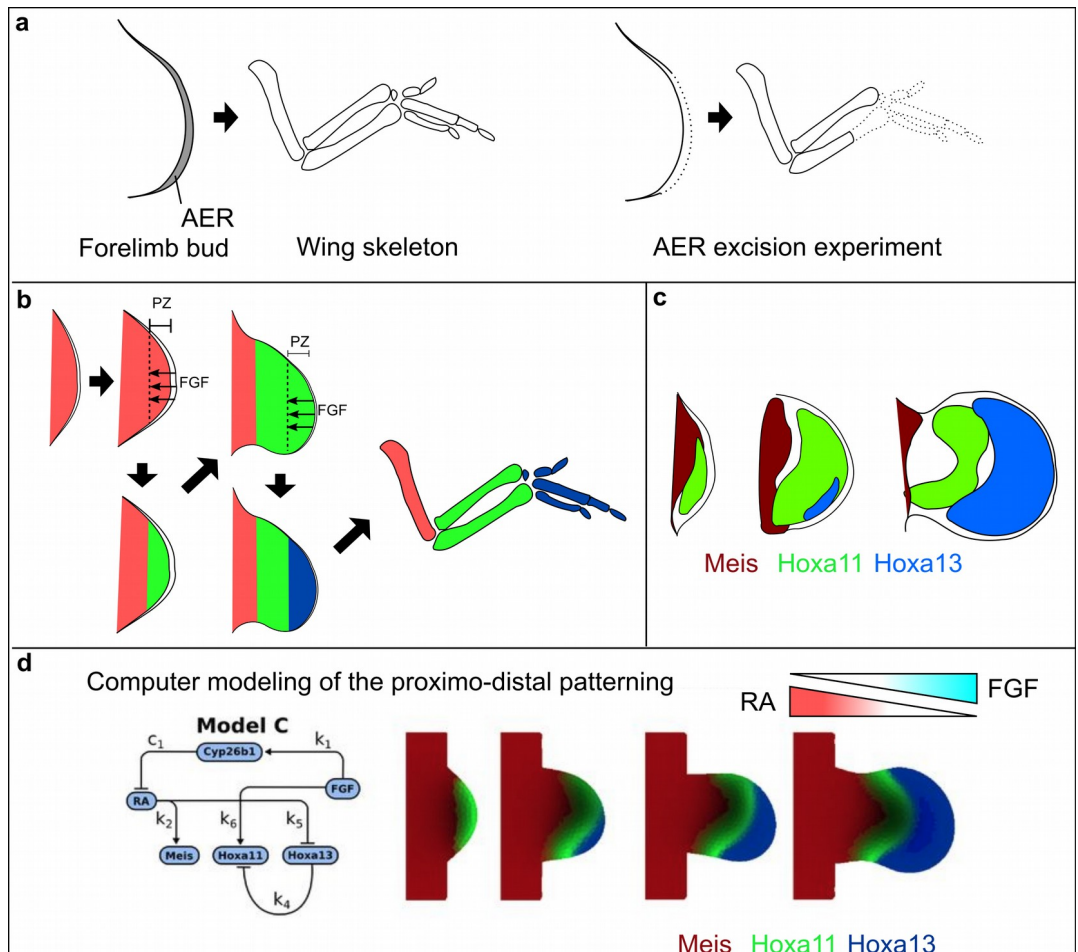

Figure 2: PD patterning in limb development. **a**, Saunders's AER experiments, adapted from Saunders 1948 [27]. B, The progress zone model, adapted from Tabin & Wolpert 2007 [30]. PZ, progress zone. C, The expression pattern of *Meis1/2, Hoxa11*, and *Hoxa13* in mouse limb buds, based on Marcader et al., 2009 [31]. D, *in silico* simulation of the two-signal model. Reproduced from Uzkudun et al., 2015 [32].

The AER is a thickened ectodermal ridge, running on the tip of limb buds, and responsible for growth and pattern formation along the proximo-distal (PD) axis. In 1948, Saunders reported his discovery that removal of the AER causes terminal limb deficiencies [27] (Fig. 2A). Based on Saunders' experiments, Wolpert suggested a "laying-down" mechanism, in which signals from the AER lays down positional information from proximal to distal in the course of growth [29]. This suggestion was later elaborated and called as the "progress zone model" [33], in which cells in a narrow domain directly below the AER (progress zone) are kept undifferentiated by AER signals, and undergo differentiation when they become far from the AER (Fig. 2B). According to this model, the positional value of cells is determined by the duration of exposure to the AER signals, which includes in the model the assumption of a clock-like mechanism within cells to measure time. In 1990s, several fibroblast growth factors (FGFs) were found to substitute the activity of the AER [34, 35], and FGFs were indeed expressed in the AER [36–38]. These series of discoveries provided the first strong evidence that morphogens controlled the pattern formation of the limb PD axis in a broad sense. Successively, alternative PD patterning models were proposed, such as the early specification model [39] and the two-signal model [30]. In the two-signal model, a gradient of FGFs from the AER and a gradient of retinoic acid from the proximal side coordinately determine positional values. Namely, FGFs positively regulate distal genes such as *Hoxa11* (a zeugopod marker) and *Hoxa13* (an autopod marker), and RA counter-acts the FGF signal to maintain proximal genes, such as *Meis1* (Fig. 2C for the expression pattern of these markers). Recently two studies provided data to show that features of both the progress zone model and the two-signal model are involved in PD patterning; FGFs are required for the activation of the distal markers, but also act to keep cells undifferentiated [40, 41], suggesting the requirement of a clock-like mechanism (see [42] for a detailed discussion). Interestingly, a data-driven computational model confirmed the validity of the two-signal model, and also indicated that the clock-like mechanism can be explained by the dynamics of gene regulatory network in the frame of the two-signal model [32] (Fig. 2D). Overall, several evidence suggest that positional information by the FGF and RA gradients control limb PD patterning.

The case of the anterio-posterior (AP) patterning is more complicated. An important signaling region for AP patterning is the ZPA, that is located in the posterior part of limb buds and can induce a mirror-imaged duplication of the limb when grafted anteriorly [28]. From this induction ability, Wolpert suggested that the ZPA provided a gradient of "polarity potential" to specify digit identity, with the highest potential close to the ZPA and the lowest potential in the anterior side of the limb bud [29] (Fig. 3A and B). More than two decades later, SHH was found to be the actual molecule responsible for AP patterning because its expression domain exactly match with the ZPA, and *Shh*-expressing cells mimic the grafting experiment of the ZPA [43]. Indirect visualization of SHH gradient with its target genes, such as *Ptch1*, also helped speculate the distribution of positional values in the limb bud [44]. In addition, *Shh* null mice form only one zeugopod element (with unclear identity) and one or no digit [45]. The result is congruent with the grafting experiment, but also indicates that SHH is not only responsible for positional information along the AP axis, but it is also required for limb growth. At this point of time, perturbations of the Shh signaling pathway linearly correlated with the initial proposal by Wolpert. However, the discovery of the phenotype of the *Shh*;*Gli3* double-knockout became difficult to reconcile with the idea that SHH provides positional information. GLI3 is known to be a main factor of the SHH pathway, and act as a transcription repressor. The repressor activity of GLI3 is inhibited by Shh, resulting in the gradient of GLI3 repressor opposite to the Shh gradient. This GLI3 repressor gradient is thought to be the actual substance of positional values in limb buds [46]. The limbs of *Gli3* mutants exhibit severe polydactyly, which was interpreted to be due to the ectopic *Shh* expression in the anterior side of the limb buds (i.e., the loss of the anterior positional value) [47]. In contrast to this interpretation, the double-knockout of GLI3 and SHH also resulted in severe polydactyly with no clear identity [48, 49], showing that GLI3 and SHH are not required for digit formation, but instead that they are likely to constrain "*the polydactylous potential of the autopod*" [49]. As we will discuss more detail in the next section, these findings are instead consistent with an alternative digit patterning mechanism based on a self-organizing Turing reaction-diffusion model [50, 51].

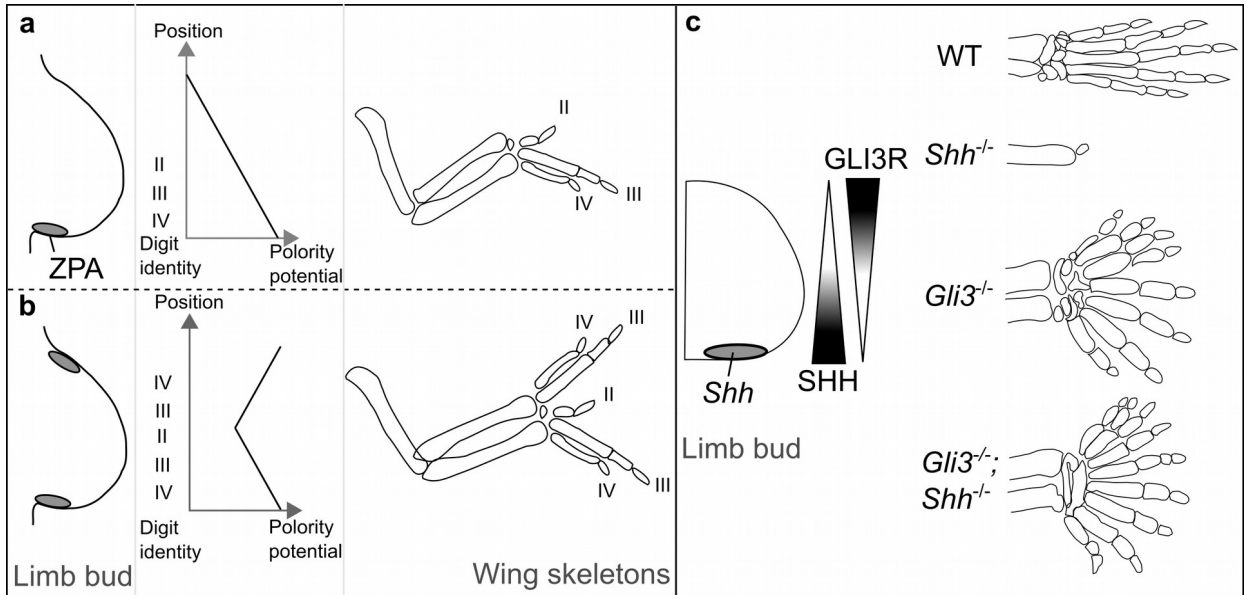

Figure 3: Positional information theory of the ZPA. **a, b,** Schemes of polority potential in a normal chick wing (**a**) and Saundars' ZPA graft experiment (**b**). Adapted from Saunders JW, Gasseling 1968 and Wolpert 1969 [28, 29]. **c,** Mouse genetics on AP patterning. Left panel, a mouse limb bud with SHH and GLI3 repressor (GLI3R) gradients, based on Riddle et al., 1993 and Wang et al., 2000 [43, 46]. Right panel, the autopods of indicated genotypes, adapted from Litingtung et al., 2002 [49].

In summary, positional information models can recapitulate some aspects of PD and AP patterning. However, in both cases they fail to explain important genetic perturbations that question the relation between the genes associated with positional information and anatomical modules. In the case of AP patterning several digits can form in the absence of a SHH gradient and only a few genes have been shown to be expressed in a specific digit. For example, *Pax9* is expressed only in the digit I and is positively regulated by GLI3 [52], but *Pax9* mutants exhibit only a preaxial polydactyly [53], instead of loss of digit I. In the case of PD patterning, loss of function of the zeugopod markers *Hoxa11* exhibit only a subtle malformation of the forelimb wrist [54]. Similarly, while *Hoxa13* and *Hoxd13* show autopod-specific expression, the double mutants of these genes exhibit truncated polydactlous limbs rather than loss of the autopod [55]. This non-linear relationship between genes and phenotypes suggests that skeletal patterning cannot be fully explained by the positional information theory and advocate for alternative patterning mechanisms as discussed in the following sections .

**Turing mechanism as local interactions**
Turing reaction-diffusion mechanism is a mathematical theory that explain spontaneous pattern formation as the results of interactions between two or more diffusive molecules. The work of Alan Turing in 1952 [2], is one of the most important contributions in theoretical biology and yet for long time a lot of doubts were raised about its validity. In this section, we briefly introduce the Turing reaction-diffusion model and present recent evidence that show its importance in developmental biology. Turing considered a set of simple reaction-diffusion equations to describe interaction between diffusible substances that he named morphogens:

$$\frac{\partial c}{\partial t} = f(c) + D \nabla^2 c \ ,$$

where *c* is the vector of molecules or morphogens, *f* represents interactions between them, and *D* is the diagonal matrix of diffusion constants. Turing showed that under specific conditions, this system could self-organize to generate periodic spatial patterns of morphogen concentration. The profound and counter-intuitive discovery of Turing is that diffusion, which usually is believe to play an equilibrating role, coupled with specific reactions could amplify random morphogen fluctuations to create spatial heterogeneity (i.e. periodic morphogen concentration patterns). The resulting patterns could be stationary periodic patterns of spots or stripes, waves or oscillations depending on parameters (Fig. 4 as an example; reviewed by [56]). This mechanisms could recapitulate various biological pattern formation such as fish skin, seashells, and the skeletal structure of limbs (see [57] for the further details about the Truing mechanism). However, the mere resemblance between simulated patterns and experimental data was not considered as a strong evidence for the Turing mechanism.

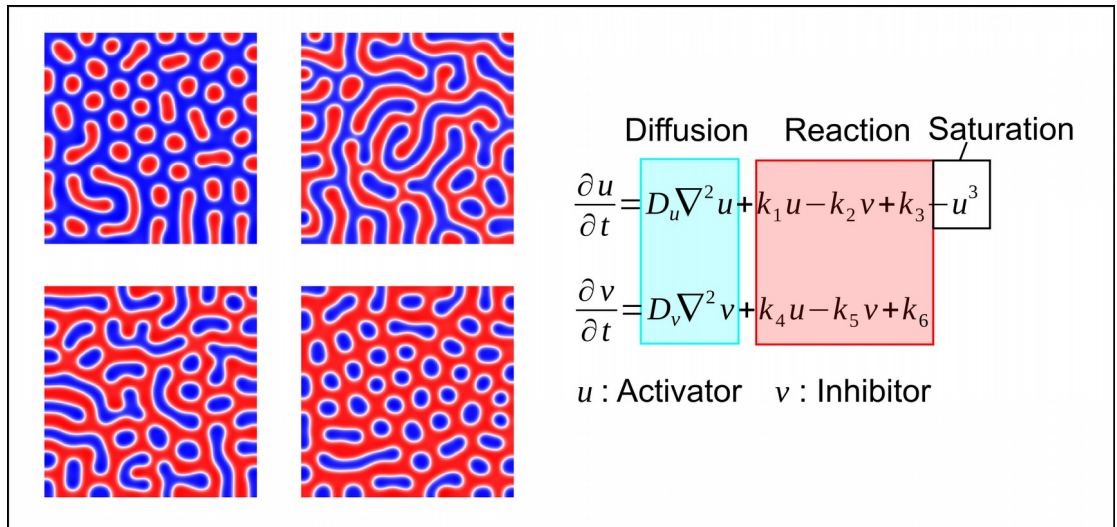

Figure 4: An example of the Turing mechanism. The top four panels show simulation results of the bottom equations with slightly different parameters, generating spot, stripe, and anti-spot patterns. The bottom equations are a linear activator-inhibitor model.

Rather, it was proposed that Turing systems required parameter fine-tuning [58], and therefore were too unreliable to control pattern formation during embryonic development [59, 60]. This assumption was based on the observation that Turing systems could undergo the diffusion-driven instability in a very narrow parameter space region [58, 61]. Therefore, a major concern in biology was that to be robust these systems required large difference in diffusivities, which was difficult to reconcile with the similar diffusion coefficients of molecules. The parameter space to obtain a Turing patterns is usually derived by performing linear instability analysis, which identifies the parameters that promote stability without diffusion and instability in the presence of diffusion. Owing to mathematical complexity, linear stability analysis was initially performed only on simple Turing systems implemented by two diffusible substances, which identified two minimal models that could generate periodic patterns: the activator-inhibitor and the substrate-depletion model. Both models required a considerable degree of differential diffusion. For example, a pervasive formulation of the activator-inhibitor model requires the inhibitor molecule to diffuse two order of magnitude faster than the activator to form a robust pattern [62], which is unrealistic for most of biological molecules. Increasing studies, however, indicate that differential diffusion is not a necessary condition for the majority of Turing models [63–65]. In particular, a recent study developed a novel method to perform an automated linear stability analysis, and demonstrated that multi-component reaction-diffusion systems that contain non-diffusive molecules expand the parameter space of diffusion-driven instability, completely eliminating the differential diffusion requirement [66].

Another argument against Turing mechanisms was the unreliability of their patterns [59]. Due to sensitivity to initial conditions, simple Turing models produce patterns that can be considered unreliable following three criteria: they can have slightly different number of stripes/spots, the orientation of these periodic patterns is random and the sequence of pattern appearance cannot be easily controlled. This unreliability has been gradually solved by adding several external controls. For example, expanding boundaries at a certain speed can reliably generate patterns with a same number of periodic elements [67]. In addition, several recent studies suggested that modulation of reaction-diffusion systems with external gradients can increase the robustness of pattern formation [68], and also control the orientation of stripes [69, 70]. However, whether or not these kinds of modulations are biologically relevant hypothesis is still being investigated.

To summarize, recent studies have revived the biological relevance of Turing's theory and have broaden its potential applications to other developmental pattering systems. Moreover, in recent years several other self-organizing patterning mechanisms based on mechanical force, cell migrations or active transport, have been proposed in addition to Turing models (see [71] for details). In the future, new self-organizing models that combine different cellular and regulatory behaviours should be explored further. Nevertheless, Turing's original model remains a seminal contribution that provides the set of theoretical conditions to explain self-organization and that could be regarded as a minimal approximation to explain pattern formation in real biological systems.

**Intertwining the Turing mechanism and positional information in limb development**
Several attempts to build the reaction-diffusion models of limb development were already made in 1970s and '80s [72–74], though strictly speaking they did not involve the Turing mechanism [75]. Most of these abstract models predicted that the different number of the skeletal elements in the stylopod, zeugopod, and autopod was controlled by the width of the limb bud along the anterio-posterior (AP) axis. This ruled out the requirement of a clock-like mechanism to determine positional identity along the PD axis. In particular, Wilby and Ede wisely discussed how their self-organizing model differed from Wolpert's positional information model by taking *talpid*$^3$ as an example [73]. *talpid*$^3$ is a chicken mutant that exhibit numerous extra digits and expanded limb buds [76]. Wilby and Ede proposed that with their reaction-diffusion model, the polydactyl *talpid*$^3$ phenotype could be readily explained by the increased width of the limb bud. In contrast, Wolpert's model was not expected to produce extra digits because the expansion of limb would not change the range of the AP morphegen gradient. The only way in which Wolpert's model could explain the appearance of new digits was if a second ZPA would be initiated in the anterior part of the limb, providing a duplication of the gradient similar to grafting experiment. Just twenty years later, without noticing Wilby and Ede's idea, the *talpid*$^3$ mutant was indeed found to have a normal *Shh* expression pattern and no duplication of the ZPA [77], which should have been supportive evidence for reaction-diffusion models. In the following decades, more sophisticated Turing mechanisms were proposed by several groups to explain limb skeletal patterning [78–84]. However, there was seemingly a tremendous gap between theoretical and experimental biologists, which was probably due to the fact that the models were rather abstract and they lacked clear experimental data to support their assumptions.

Although the Turing mechanism was not a mainstream of investigation for long time, it was gradually realized that the patterning mechanism of limb development could not be understood only on the basis of the positional information theory [51]. As mentioned in the previous section, the critical turning point was the surprising discovery of the polydactylous phenotype of *Gli3;Shh* double mutants [48, 49], in which the signaling center connected to AP patterning by positional information was completely abolished. These mutants suggested that digits formed spontaneously by a self-organizing mechanism, that produced more digits in larger limbs. Almost a decade later, three studies confirmed this hypothesis [69, 85, 86]. These study showed that not only more digits were created in expanded limbs that lacked GLI3, but that in this genetic background, the number of digits could be further increased by removing *Hoxd* and *Hoxa* genes. These genes were originally thought to encode positional values for digit specification. However, in [69], a quantitative comparison between combinatorial mutations of *Hoxa* and *Hoxd* genes and an abstract Turing model showed that *Hox* genes act as negative regulators of the wavelength of of a Turing mechanism responsible for periodic digit patterning. Interestingly, the authors went further to speculate that the pentadactyl limb evolved by a gradual increase in the dose of distal *Hox* expression to reduce the numerous skeletal elements of ancestral fish fins. As discussed later, this speculation seems to agree with experimental observations of catshark fin buds. This work provided a new interpretation of the function of *Hox* genes and a mechanical explanation of polydactylous phenotypes in the absence of positional information. This was also the first study to provide evidence that Turing models combined external modulations such as *Hox* gene dose and gradients from the AER, thereby robustly reproducing digit patterning.

The molecular candidates responsible for the Turing mechanism (Turing molecules), however, were still unknown. A previous study proposed *Tgfb2* as a strong Turing molecule candidate, because it enhances chondrogenic differentiation and it is localized with prechondrogenic condensation in micromass cultures [84]. However, the *Tgfb2* expression in limb development starts after the digit pattern appears [87]. In addition, *Tgfb2* null mutants show only a minor patterning defect [88]. FGFs are also often mentioned as a candidate [79], but there is no evidence for the co-localization of their signal with condensations. Although an FGF receptor, *Fgfr2* (more precisely its IIIc isoform) is expressed in the prechondrogenic mesenchyme [89, 90], the loss of *Fgfr2* in the limb mesenchyme affects only a minor growth defect [91]. Similarly, no patterning defects are also observed in the limbs lacking galectins (e.g., [92]), which were proposed as Turing molecules to explain digit patterning in chick embryos.

Strong experimental evidence in support of specific Turing molecules was provided only more recently [87]. This study proposed that Turing molecules should be spatially distributed either in-phase or out-of-phase periodic patterns with the digits, reflecting the distributions observed in the two minimal Turing models, the activator-inibitor and subtrate-depletion model. Moreover, it proposed that candidates should be active as soon as the digit patterning process is occurring. By performing a detailed spatio-temporal analysis of candidates with comparative microarray analysis, *in situ* hybridization and analysis of protein distributions, this study identified three promising molecules as Turing candidates: BMP, SOX9, and WNT (referred to as the BSW model; Fig. 5A). SOX9 is the earliest chondrogenic marker (therefore in-phase with digit primordia), and is required for prechondrogenic condensation [93]. This study showed that *Bmp2* expression and WNT signal exhibited out-of-phase distributions with respect to Sox9, that was congruent with the requirement for Turing molecules. In addition, perturbations of the BMP and WNT pathways are consisted with their roles as Turing molecules that regulate Sox9. It is worth emphasizing that in this study, three technical advances played a central role to test the molecular candidates. Firstly, the development of a computational model based accurate quantification of 2D limb bud morphology, including information about growth, tissue movement, and spatial gene expression dynamics [94]. This allowed the authors to compare realistic simulations of digit patterning with experimental expression patterns. Secondly, the use of an *ex utero* limb culture method, which utilizes the air liquid interface cell culture system, to perform perturbation and test model predictions in a near *in vivo* environment. Thirdly, the development of a novel Turing model with three components that include a non-diffusing reactant corresponding to SOX9. This last point was a major difference with previous models, which considered only two diffusible substances. Multi-component models with non-diffusible elements relax the parameter constraint required to form Turing patterns and increase the robustness of patter formation. This extended Turing model, known as the BSW model, captures the essential dynamics of skeletal pattern formation showed by *Sox9* expression patterns. In addition, the model integrates a Turing network with positional signals, such as FGFs and HOX genes, that helps to generate the properly aligned digit primordia in a reproducible manner. The theoretical idea that external gradients could modulate Turing systems to obtain specific patterns was already proposed previously [62]. However, the BSW model represents the first biologically relevant realization of this idea supported by experimental data. Taken together, this study does not only represent a major step forward to understand limb development, but it also invokes a conceptual innovation by showing that Turing mechanisms and positional information can act simultaneously, and should not be considered as alternative mechanisms [95, 96].

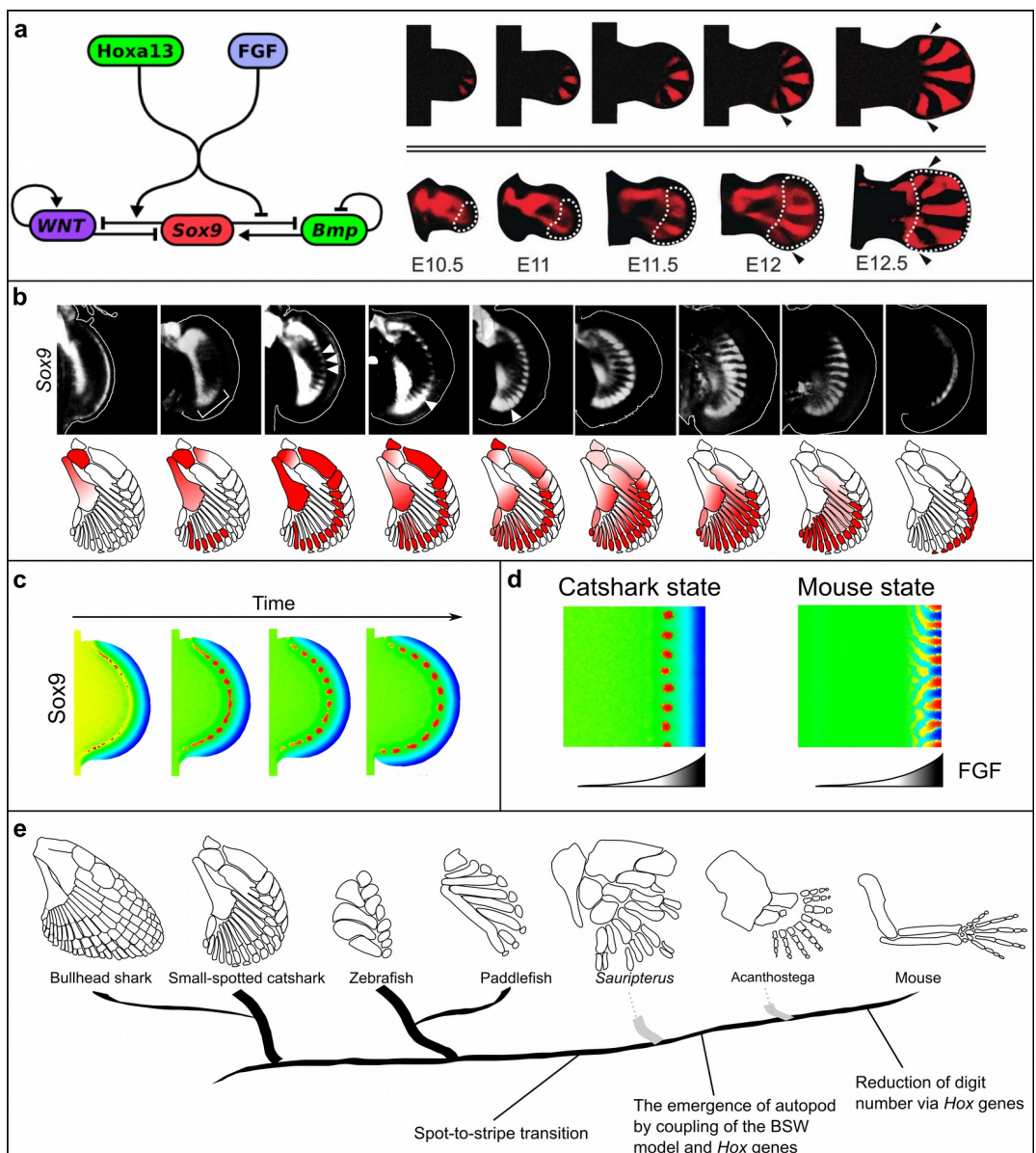

Figure 5: Computer modeling of limb and fin development. **a**, The BSW network (left; the scheme was kindly provided by James Sharpe) and comparison between a simulation result (top right) and real *Sox9* expression (right bottom). From Raspopovic et al., SCIENCE 345:566 (2014) [87]. Reprinted with permission from AAAS. **b**, Time-series of *Sox9* expression of the catshark pectoral fin buds. Bottoms are corresponding skeletal elements. **c**, A simulation of *Sox9* expression

of the catshark fin buds. **d**, The difference between fins and limbs. B-D are reproduced from Onimaru et al., 2016 [97]. E, The diverse morphology of fins. Note that distal nodular radials are widely recognized in fish fins, but *Sauripterus* (a fossil fish) may have had different type of fins. A autopod-like structure is seen in *Acanthostega* (a stem fossil tetrapod), but it shows polydactyly. *Sauripterous* and *Acanthostega* are adapted from Davis et al., 2004 and Caotes, 1996, respectively [98, 99].

Indeed, positional information and the Turing mechanism have often been seen as contrasting patterning strategies. The former has a hierarchical nature, because morphogen gradients are upstream and genes for differentiation are downstream. Moreover, it requires a signaling center that imposes a global asymmetry, where identity is uniquely determined as a function of distance from the signaling source. Turing models on the contrary are self-organizing, and are based on local molecular interactions that amplify random fluctuations to generate a periodic pattern. This model generate identical elements that are regularly distributed in space. Positional information and the Turing mechanism, however, can act together to form intermediate structures, e.g., periodic pattern with slightly different wavelength [95]. In such intermediate systems, the identity of the resulting anatomical modules can be vague and quantitative rather than qualitative. As seen by *Gli3* and *Hox* gene mutants, the loss of genes related to positional information by genetic perturbations may accentuate the nature of the Turing mechanism.

**Emergence of homology by coupling Turing mechanism and positional information**
As we have seen in the previous sections, an integrative approach is required to understand the evolution of fins and limbs. We covered a wide range of topics such as classical comparative anatomy, developmental biology, and mathematical models. However, the above discussion can be summarized into two main points: 1. the homology problem between fins and limbs is likely due to poor understandings of the developmental mechanism that underlies skeletal pattern formation; 2. Turing mechanism and positional information control together limb development. In this section, we discuss how both mechanisms are also involved in the development of fish fins.

Molecular studies of the fin-to-limb transition have suffered from similar problems to the ones faced inlimb development that we mentioned in the previous section. Although key positional information genes, such as HOX genes, SHH, and FGFs, were analyzed to investigate fin development in several species including sharks, rays and ray-finned fishes, the observed differences were not readily interpretable. For example, the overlapping expression domains of *Hoxa11* and *Hoxa13* seem to be a common feature in pectoral fin buds, while in tetrapod limbs, *Hoxa11* and *Hoxa13* are expressed in the zeugopod and autopod domains separately [100, 101]. However, as discussed earlier, current data shows that these genes alone do not affect skeletal patterning process in mouse limb buds, therefore this regulatory difference alone cannot fully explain the anatomical difference between fins and limbs. Secondly, a study [26] showed that the gross expression domains of genes related to the AP positional information (or SHH targets) are shifted to the posterior side of catshark pectoral fin buds. Experimentally shifting of these expression domains anteriorly caused a fusion of the anterior radials to metapterygium, which is correlated with the pro- and mesopterygium loss during the fin-to-limb transition. However, owing to the lack of the one-to-one correspondence between gene expression domains and skeletal elements, it remains unclear how these genes regulate skeletal patterning.

Fish fins exhibit a greater variety of skeletal patterns than tetrapod limbs (Fig. 5E). Most fins seem to have an affinity to the basic characteristic of the Turing pattern, because they are composed of numerous and regular periodic elements, but also show a weak asymmetric trend along both of the AP and PD axes. Therefore, given the aforementioned nature of limb development, the pattern formation of fish fins strongly points to a Turing mechanism that is still weakly controlled by positional information. This idea was recently tested in [97], where a combination of data-driven *in silico* modeling and *in vivo* experiments was used to investigate catshark pectoral fin development. This study showed that the Turing mechanism driven by the interactions of BMP, SOX9, and WNT seems to be conserved at least between mouse digits and the distal fin elements of the catshark (Fig. 5B for *Sox9* expression data and 5C for simulated *Sox9* expression with the BSW model). However, there are several differences that are likely responsible for their different anatomical structures: 1. in the catshark, the distal fin radials arise from a periodic spot pattern of *Sox9* expression (Fig. 5B), in contrast to the stripe pattern in mouse digit formation, highlighting one of the major parameters that cause the anatomical difference between fins and limbs. 2. In catshark fins, the BSW model seems to be decoupled from the distal Hox genes (yet still regulated by FGFs). *Sox9* expression in catshark pectoral fin buds is located proximal to the distal *Hoxa13* expression without an overlapping domain, which is congruent with the reverse correlation between the amount of *Hox* genes and the number of skeletal elements [69]. Interestingly, two independent groups concurrently suggested that HOX13s in fish fins regulate the differentiation of fin rays, but not radials [22, 23]. In addition, the computational model predicts that modulation of WNT-related parameters shifts the *Sox9* expression domain distally, and changes the spot pattern into a stripe pattern (Fig. 5D). Therefore, it is tempting to hypothesize that a distal shift of *Sox9* expression and a pattern change from spots to stripes were critical events prior to the *de novo* acquisition of the digits. Subsequently, the acquisition of wavelength regulation by the distal Hox genes [69] may have resulted in the emergence of the homologous autopod and a gradual reduction of the number of digits until the current pentadactyl state (Fig. 5E).

It is worth noting that there is an earlier attempt to simulate the diverse morphologies of fins with the Turing mechanism [80]. In this study, a single computational framework simulated overall skeletal patterns of fins and limbs of various species including the catshark, obtaining a good degree of resemblance between known skeletal patterns and simulations. The models developed in this study however were not constrained by experimental data. Instead they arbitrarily modulated the AP width and wavelength without comparison with real data, which limit the predictive power of the model. In fact, the *Sox9* expression pattern of catshark fin buds [97] does not fit with the catshark simulation presented in this study. This highlights that to draw a biologically meaningful conclusion, computational models should be constrained with real data.

In this section, we discussed how the underling developmental patterning mechanisms plays a central role to understand the morphological difference between fins and limb. Evidence indicates that a Turing mechanism modulated by positional information is very likely to underlie skeletal pattern formation in both fins and limbs. The diverse morphologies of fins radials and digits can be traced to changes in the modulations of a Turing mechanism to generate topologically different patterns, such as spots and stripes with slight changes in parameters. This is a major conceptual shift to understand the fin-to-limb evolution, and it provides a completely new framework to infer homology and similarity between appendages. Nevertheless, there are many aspects of the fin-to-limb transition remain unclear. For example, the biological significance of the metapterygium is not yet solved. *Sox9* expression data (Fig. 5B; [97]) suggest that the metapterygium at least in the catshark, is not formed by a branching process that can be inferred from the adult skeletal pattern. Instead, the distal elements appear separately from the proximal ones and they are connected later. Therefore, branching-like pattern, which is thought to be the characteristic of the metapterygium, may not represent a common developmental mechanism derived from the last common ancestor of jawed vertebrates. Another untouched issue is the actual genetic changes that are responsible for the different dynamics of fin and limb pattern formation. Because several shark genome sequences were recently released [102], comparative genome analysis is now possible and will help us draw a more detail genotype-phenotype map of fins and limbs.

**Evolutionary systems biology toward the genotype-phenotype mapping**
From all the above discussions, we draw three general conclusions:
> 1) Anatomical homology can be recognized when the elements of an organ exhibit global asymmetry induced by morphogen gradients.
> 2) The periodic nature of self-organizing mechanisms obscures homologous relationship of elements.
> 3) Because global asymmetries and self-organizing mechanisms can work as a single system, there could be intermediate states where homologous relationship becomes vague.

These conclusions may not be applicable to every case of evolutionary studies, but they will work as precaution by showing that some homology problems may not be solvable because of the nature of developmental mechanisms. The avian digit identity problem—a debate if avian digits should be counted as "I, II, III" or "II, III, IV"—might be the case of such "overdiagnosis". Interestingly, evolutionary drifts between periodic and identity-containing patterns appear to be observed in other systems, such as body segmentation in insects and computational experiments [103–105]. Therefore, even if a morphological part was derived from a common ancestor, the homologous relation can disappear by changes in the developmental mechanisms responsible for the identity.

We highlighted efforts to understand the evolution of multicellular systems by combining computational methods, mathematical theories, and experimental biology. Such integral studies are necessary to overcome the counter-intuitive nature of gene regulatory dynamics. One of the critical factors that we did not cover is genome information. Currently, it is not possible to infer gene regulatory networks and model parameters from genome sequences. Because evolution is driven by genetic mutations, bottom-up approaches to construct gene regulatory models from genome information are needed to draw the whole picture of the genotype-phenotype map. The recent advancement in DNA-sequence technologies has improved the availability of genome sequences from various species. However, currently, owing to the complicated organization of the genome, the ability to extract functional information such as gene regulatory networks from genome sequences remains limited. Several recent studies indicate that deep learning–based methods are promising to interpret genome information (e.g., [106–108]). Developing methods to understand what is encoded in genomes will be required to close the gap between genotypes and phenotypes in the future.

In conclusion, we have discussed how the nature of morphological evolution depends on developmental mechanisms. Because of the counter-intuitive dynamics of gene interactions, combinatorial approaches of computer modeling, theoretical and experimental biology is essential to understand the evolution of multicellular organisms. Therefore, finding a good collaboration between researchers from different fields or educational opportunities to learn different fields will be a key to stimulate the growth of evolutionary systems biology.

**Acknowledgements**
We thank Dr. Tatsuya Hirasawa and Dr. Xavier Diego for fruitful discussions; Dr. James Sharpe for critical comments on the manuscript. This work was supported in part by JSPS KAKENHI grant number 17K15132, a Special Postdoctoral Researcher Program of RIKEN, and a research grant from MEXT to RIKEN Center for Biosystems Dynamics Research to K.O., and by the María de Maeztu Unit of Excellence MDM-2016-0687 (MINECO, Spain) to L.M..

**References**

1. Alberch P (1991) From genes to phenotype: dynamical systems and evolvability. Genetica

2. Turing AM (1952) The Chemical Basis of Morphogenesis. Philos Trans R Soc B, 237:37–72

3. Wagner GP (2014) Homology, genes and evolutionary innovation. Princeton university press

4. Owen R (1848) On the archetype and homologies of the vertebrate skeleton. John van Voorst , Paternoster Row., London

5. Mayr E (1982) The growth of biological thought: Diversity, Evolution, and Inheritance

6. Shubin NH, Alberch P (1986) A morphogenetic approach to the origin and basic organization of the tetrapod limb. Evol Biol 20:318–390

7. Gegenbaur C (1865) Untersuchungen zur vergleichenden anatomie der wirbeltiere. Vol. II. Wilhelm Engelmann, Leipzig

8. Jarvik E (1980) Basic structure and evolution of vertebrates. Vol. 2. Academic Press Inc. (London) LTD., London

9. Watson DMS (1913) On the primitive tetrapod limb. Anat Anz 44:24–27

10. Holmgren N (1952) An embryological analysis of the mammalian carpus and its bearing upon the question of the origin of the tetrapod limb. Acta Zool. 33:1–115

11. Tarchini B, Duboule D (2006) Control of Hoxd genes' collinearity during early limb development. Dev Cell 10:93–103. https://doi.org/10.1016/j.devcel.2005.11.014

12. Wagner GP, Larsson HCE (2007) Fins and limbs in the study of evolutionary novelties. In: Hall BK (ed) Fins into Limbs: Evolution, Development, and Transformation. The University of Chicago Press, pp 49–61

13. Cohn MJ, Lovejoy CO, Wolpert L, Coates MI (2002) Branching, segmentation and the metapterygial axis: Pattern versus process in the vertebrate limb. BioEssays 24:460–465. https://doi.org/10.1002/bies.10088

14. Nelson CE, Morgan B a, Burke AC, et al (1996) Analysis of Hox gene expression in the chick limb bud. Development 122:1449–1466. https://doi.org/10.1038/342767a0

15. Dolle P, Izpisua-Belmonte, J. C., Falkenstein H, Renucci A, et al (1989) Coordinate expression of the murine Hox-5 complex homoeobox-containing genes during limb pattern formation. Nature 342:767–772

16. Shubin N, Tabin C, Carroll S (1997) Fossils, genes and the evolution of animal limbs. Nature 388:639–648

17. Sordino P, van der Hoeven F, Duboule D (1995) Hox gene expression in teleost fins and the origin of vertebrate digits. Nature 375:678–681

18. Davis MC, Dahn RD, Shubin NH (2007) An autopodial-like pattern of Hox expression in the fins of a basal actinopterygian fish. Nature 447:473–476. https://doi.org/10.1038/nature05838

19. Freitas R, Zhang G, Cohn MJ (2007) Biphasic Hoxd gene expression in shark paired fins reveals an ancient origin of the distal limb domain. PLoS One 2:e754

20. Tulenko FJ, Augustus GJ, Massey JL, et al (2016) HoxD expression in the fin-fold compartment of basal gnathostomes and implications for paired appendage evolution. Sci Rep 6:. https://doi.org/10.1038/srep22720

21. Ahn D, Ho RK (2008) Tri-phasic expression of posterior Hox genes during development of pectoral fins in zebrafish: implications for the evolution of vertebrate paired appendages. Dev Biol 322:220–233

22. Nakamura T, Gehrke AR, Lemberg J, et al (2016) Digits and fin rays share common developmental histories. Nature 537:225–228. https://doi.org/10.1038/nature19322

23. Tulenko FJ, Massey JL, Holmquist E, et al (2017) Fin-fold development in paddlefish and catshark and implications for the evolution of the autopod. Proc R Soc B Biol Sci 284:20162780. https://doi.org/10.1098/rspb.2016.2780

24. Freitas R, Gómez-Marín C, Wilson JM, et al (2012) Hoxd13 Contribution to the Evolution of Vertebrate Appendages. Dev Cell 23:1219–1229. https://doi.org/10.1016/j.devcel.2012.10.015

25. Clack JA (2009) The fin to limb transition: new data, interpretations, and hypotheses from paleontology and developmental biology. Annu Rev Earth Planet Sci 37:163–179. https://doi.org/10.1146/annurev.earth.36.031207.124146

26. Onimaru K, Kuraku S, Takagi W, et al (2015) A shift in anterior–posterior positional information underlies the fin-to-limb evolution. Elife 4:e07048. https://doi.org/10.7554/eLife.07048.001

27. Saunders JW (1948) The proximo-distal sequence of origin of the parts of the chick wing and the role of the ectoderm. J Exp Zool 108:363–403. https://doi.org/10.1002/jez.1401080304

28. Saunders JW, Gasseling MT (1968) Ectodermal-mesenchymal interactions in the origin of limb symmetry. In: Epithelial Mesenchymal Interactions. pp 78–97

29. Wolpert L (1969) Positional information and the spatial pattern of cellular differentiation. J Theor Biol 25:1–47

30. Tabin C, Wolpert L (2007) Rethinking the proximodistal axis of the vertebrate limb in the molecular era. Genes Dev 21:1433–1442. https://doi.org/10.1101/gad.1547407

31. Mercader N, Selleri L, Criado LM, et al (2009) Ectopic Meis1 expression in the mouse limb bud alters P-D patterning in a Pbx1-independent manner. Int J Dev Biol 53:1483–1494

32. Uzkudun M, Marcon L, Sharpe J (2015) Data-driven modelling of a gene regulatory network for cell fate decisions in the growing limb bud. Mol Syst Biol 11:815–815. https://doi.org/10.15252/msb.20145882

33. Summerbell D, Lewis JH, Wolpert L (1973) Positional Information in chick limb morphogenesis. Nature 244:492–496. https://doi.org/10.1038/244492a0

34. Fallon JF, López A, Ros MA, et al (1994) FGF-2: Apical ectodermal ridge growth signal for chick limb development. Science 264:104–107. https://doi.org/10.1126/science.7908145

35. Niswander L, Tickle C, Vogel A, et al (1993) FGF-4 replaces the apical ectodermal ridge and directs outgrowth and patterning of the limb. Cell 75:579–587. https://doi.org/10.1016/0092-8674(93)90391-3

36. Niswander L, Martin GR (1992) Fgf-4 expression during gastrulation, myogenesis, limb and tooth development in the mouse. Development 114:755–768

37. Ohuchi H, Yoshioka H, Tanaka A, et al (1994) Involvement of androgen-induced growth factor (FGF-8) gene in mouse embryogenesis and morphogenesis. Biochem Biophys Res Commun 204:882–888. https://doi.org/10.1006/bbrc.1994.2542

38. Heikinheimo M, Lawshé A, Shackleford GM, et al (1994) Fgf-8 expression in the post-gastrulation mouse suggests roles in the development of the face, limbs and central nervous system. Mech Dev 48:129–138. https://doi.org/10.1016/0925-4773(94)90022-1

39. Dudley AT, Ros MA, Tabin CJ (2002) A re-examination of proximodistal patterning during vertebrate limb development. Nature 418:539–544. https://doi.org/10.1038/nature00945

40. Cooper KL, Hu JK-H, ten Berge D, et al (2011) Initiation of proximal-distal patterning in the vertebrate limb by signals and growth. Science 332:1083–1086

41. Roselló-Díez A, Ros M a, Torres M (2011) Diffusible signals, not autonomous mechanisms, determine the main proximodistal limb subdivision. Science 332:1086–1088. https://doi.org/10.1126/science.1199489

42. Delgado I, Torres M (2016) Gradients, waves and timers, an overview of limb patterning models. In: Seminars in Cell and Developmental Biology. Elsevier, pp 109–115

43. Riddle RD, Johnson RL, Laufer E, Tabin C (1993) Sonic hedgehog mediates the polarizing activity of the ZPA. Cell 75:1401–1416. https://doi.org/10.1016/0092-8674(93)90626-2

44. Marigo V, Scott MP, Johnson RL, et al (1996) Conservation in hedgehog signaling: induction of a chicken patched homolog by Sonic hedgehog in the developing limb. Development 122:1225–1233

45. Chiang C, Litingtung Y, Harris MP, et al (2001) Manifestation of the limb prepattern: limb development in the absence of sonic hedgehog function. Dev Biol 236:421–435. https://doi.org/10.1006/dbio.2001.0346

46. Wang B, Fallon J, Beachy P (2000) Hedgehog-regulated processing of Gli3 produces an anterior/posterior repressor gradient in the developing vertebrate limb. Cell 100:423–434. https://doi.org/10.1016/S0092-8674(00)80678-9

47. Masuya H, Sagai T, Moriwaki K, Shiroishi T (1997) Multigenic control of the localization of the zone of polarizing activity in limb morphogenesis in the mouse. Dev Biol 182:42–51

48. Te Welscher P, Zuniga A, Kuijper S, et al (2002) Progression of vertebrate limb development through SHH-mediated counteraction of GLI3. Science 298:827–830. https://doi.org/10.1126/science.1075620

49. Litingtung Y, Li Y, Fallon JF, Chiang C (2002) Shh and Gli3 are dispensable for limb skeleton formation but regulate digit number and identity. Nature 418:979–983. https://doi.org/10.1038/nature01033

50. Newman SA (2007) The Turing mechanism in vertebrate limb patterning. Nat Rev Mol Cell Biol 8:. https://doi.org/10.1038/nrm1830-c1

51. Tickle C (2006) Making digit patterns in the vertebrate limb. Nat. Rev. Mol. Cell Biol. 7:45–53

52. McGlinn E, Van Bueren KL, Fiorenza S, et al (2005) Pax9 and Jagged1 act downstream of Gli3 in vertebrate limb development. Mech Dev 122:1218–1233. https://doi.org/doi:10.1016/j.mod.2005.06.012

53. Peters H, Neubuser A, Kratochwil K, Balling R (1998) Pax9-deficient mice lack pharyngeal pouch derivatives and teeth and exhibit craniofacial and limb abnormalities. Genes Dev 49:2735–2747

54. Small KM, Potter SS (1993) Homeotic transformations and limb defects in Hox A11 mutant mice. Genes Dev 7:2318–2328. https://doi.org/10.1101/gad.7.12a.2318

55. Fromental-Ramain C, Warot X, Messadecq N, et al (1996) Hoxa-13 and Hoxd-13 play a crucial role in the patterning of the limb autopod. Development 122:2997–3011

56. Kondo S, Miura T (2010) Reaction-diffusion model as a framework for understanding biological pattern formation. Science 329:1616–1620

57. Murray JD (2003) Mathematical Biology II - Spatial Models and Biomedical Applications, 3rd ed. Springer-Verlag, New York

58. Murray JD (1982) Parameter space for turing instability in reaction diffusion mechanisms: A comparison of models. J Theor Biol 98:143–163. https://doi.org/10.1016/0022-5193(82)90063-7

59. Bard J, Lauder I (1974) How well does Turing's theory of morphogenesis work? J Theor Biol 45:501–531. https://doi.org/10.1016/0022-5193(74)90128-3

60. Maini PK, Woolley TE, Baker RE, et al (2012) Turing's model for biological pattern formation and the robustness problem. Interface Focus 2:487–496

61. Butler T, Goldenfeld N (2011) Fluctuation-driven Turing patterns. Phys Rev E - Stat Nonlinear, Soft Matter Phys 84:. https://doi.org/10.1103/PhysRevE.84.011112

62. Gierer A, Meinhardt H (1972) A theory of biological pattern formation. Kybernetik 12:30–39. https://doi.org/10.1007/BF00289234

63. White KAJ, Gilligan CA (1998) Spatial Heterogeneity in Three-Species, Plant-Parasite-Hyperparasite, Systems. PhilTrans R Soc Lond B 353:543–557

64. Korvasová K, Gaffney EA, Maini PK, et al (2015) Investigating the Turing conditions for diffusion-driven instability in the presence of a binding immobile substrate. J Theor Biol. https://doi.org/10.1016/j.jtbi.2014.11.024

65. Marcon L, Diego X, Sharpe J, Müller P (2016) High-throughput mathematical analysis identifies turing networks for patterning with equally diffusing signals. Elife 5:. https://doi.org/10.7554/eLife.14022

66. Diego X, Marcon L, Müller P, Sharpe J (2018) Key Features of Turing Systems are Determined Purely by Network Topology. Phys Rev X 8:. https://doi.org/10.1103/PhysRevX.8.021071

67. Crampin EJ, Gaffney EA, Maini PK (1999) Reaction and diffusion on growing domains: Scenarios for robust pattern formation. Bull Math Biol 61:1093–1120. https://doi.org/10.1006/bulm.1999.0131

68. Pecze L (2018) A solution to the problem of proper segment positioning in the course of digit formation. BioSystems. https://doi.org/10.1016/j.biosystems.2018.04.005

69. Sheth R, Marcon L, Bastida MF, et al (2012) Hox genes regulate digit patterning by controlling the wavelength of a Turing-type mechanism. Science 338:1476–1480. https://doi.org/10.1126/science.1226804

70. Hiscock TW, Megason SG (2015) Orientation of Turing-like Patterns by Morphogen Gradients and Tissue Anisotropies. Cell Syst 1:408–416. https://doi.org/10.1016/j.cels.2015.12.001

71. Hiscock TW, Megason SG (2015) Mathematically guided approaches to distinguish models of periodic patterning. Development 142:409–419. https://doi.org/10.1242/dev.107441

72. Goodwin BC, Trainor LEH (1983) The ontogeny and phylogeny of the pendactyl limb. Dev Evol 75–98

73. Wilby OK, Ede DA (1975) A model generating the pattern of cartilage skeletal elements in the embryonic chick limb. J Theor Biol 52:199–217. https://doi.org/https://doi.org/10.1016/0022-5193(75)90051-X

74. Newman SA, Frisch HL (1979) Dynamics of skeletal pattern formation in developing chick limb. Science 205:662–668

75. Othmer HG (1986) On the Newman-Frisch model of limb chondrogenesis. J Theor Biol 121:505–508. https://doi.org/10.1016/S0022-5193(86)80105-9

76. Ede DA, Kelly WA (1964) Developmental abnormalities in the trunk and limbs of the <em>talpid</em><sup>3</sup> mutant of the fowl. J Embryol Exp Morphol 12:339 LP – 356

77. Francis-West PH, Robertson KE, Ede DA, et al (1995) Expression of genes encoding bone morphogenetic proteins and sonic hedgehog in talpid (ta3) limb buds: Their relationships in the signalling cascade involved in limb patterning. Dev Dyn 203:187–197. https://doi.org/10.1002/aja.1002030207

78. Chaturvedi R, Huang C, Kazmierczak B, et al (2005) On multiscale approaches to three-dimensional modelling of morphogenesis. J R Soc Interface 2:237–253. https://doi.org/10.1098/rsif.2005.0033

79. Hentschel HGE, Glimm T, Glazier JA, Newman SA (2004) Dynamical mechanisms for skeletal pattern formation in the vertebrate limb. Proc R Soc B Biol Sci 271:1713–1722. https://doi.org/10.1098/rspb.2004.2772

80. Zhu J, Zhang YT, Alber MS, Newman SA (2010) Bare bones pattern formation: A core regulatory network in varying geometries reproduces major features of vertebrate limb development and evolution. PLoS One 5:e10892. https://doi.org/10.1371/journal.pone.0010892

81. Miura T, Shiota K, Morriss-Kay G, Maini PK (2006) Mixed-mode pattern in Doublefoot mutant mouse limb-Turing reaction-diffusion model on a growing domain during limb development. J Theor Biol 240:562–573

82. Miura T, Shiota K (2000) Extracellular matrix environment influences chondrogenic pattern formation in limb bud micromass culture: Experimental verification of theoretical models. Anat Rec 258:100–107. https://doi.org/10.1002/(SICI)1097-0185(20000101)258:1<100::AID-AR11>3.0.CO;2-3

83. Miura T, Maini PK (2004) Speed of pattern appearance in reaction-diffusion models: Implications in the pattern formation of limb bud mesenchyme cells. Bull Math Biol 66:627–649. https://doi.org/10.1016/j.bulm.2003.09.009

84. Miura T, Shiota K (2000) TGFβ2 acts as an "activator" molecule in reaction-diffusion model and is involved in cell sorting phenomenon in mouse limb micromass culture. Dev Dyn 217:241–249. https://doi.org/10.1002/(SICI)1097-0177(200003)217:3<241::AID-DVDY2>3.0.CO;2-K

85. Sheth R, Grégoire D, Dumouchel A, et al (2013) Decoupling the function of Hox and Shh in developing limb reveals multiple inputs of Hox genes on limb growth. Development 140:2130–2138. https://doi.org/10.1242/dev.089409

86. Sheth R, Bastida MF, Ros M (2007) Hoxd and Gli3 interactions modulate digit number in the amniote limb. Dev Biol 310:430–441. https://doi.org/10.1016/j.ydbio.2007.07.023

87. Raspopovic J, Marcon L, Russo L, Sharpe J (2014) Digit patterning is controlled by a Bmp-Sox9-Wnt Turing network modulated by morphogen gradients. Science 345:566–70

88. Sanford LP, Ormsby I, Groot AC, et al (1997) TGFβ2 knockout mice have multiple developmental defects that are non-overlapping with other TGFβ knockout phenotypes. Development 124:2659–2670. https://doi.org/10.1086/597422.Tumor

89. Szebenyi G, Savage MP, Olwin BB, Fallon JF (1995) Changes in the expression of fibroblast growth factor receptors mark distinct stages of chondrogenesis in vitro and during chick limb skeletal patterning. Dev Dyn 204:446–456. https://doi.org/10.1002/aja.1002040410

90. Sheeba CJ, Andrade RP, Duprez D, Palmeirim I (2010) Comprehensive analysis of fibroblast growth factor receptor expression patterns during chick forelimb development. Int J Dev Biol 54:1515–1524. https://doi.org/10.1387/ijdb.092887cs

91. Eswarakumar VP, Monsonego-Ornan E, Pines M, et al (2002) The IIIc alternative of Fgfr2 is a positive regulator of bone formation. Development 129:3783–3793

92. Georgiadis V, Stewart HJS, Pollard HJ, et al (2007) Lack of galectin-1 results in defects in myoblast fusion and muscle regeneration. Dev Dyn 236:1014–1024. https://doi.org/10.1002/dvdy.21123

93. Liu C-F, Angelozzi M, Haseeb A, Lefebvre V (2018) SOX9 is dispensable for the initiation of epigenetic remodeling and the activation of marker genes at the onset of chondrogenesis. Development 145:dev164459. https://doi.org/10.1242/dev.164459

94. Marcon L, Arqués CG, Torres MS, Sharpe J (2011) A computational clonal analysis of the developing mouse limb bud. PLoS Comp Biol 7:e1001071

95. Green JBA, Sharpe J (2015) Positional information and reaction-diffusion: two big ideas in developmental biology combine. Development 142:1203–1211. https://doi.org/10.1242/dev.114991

96. Miura T (2013) Turing and wolpert work together during limb development. Sci Signal 6:. https://doi.org/10.1126/scisignal.2004038

97. Onimaru K, Marcon L, Musy M, et al (2016) The fin-to-limb transition as the re-organization of a Turing pattern. Nat Commun 7:11582. https://doi.org/10.1038/ncomms11582

98. Davis MC, Shubin NH, Daeschler EB (2004) A new specimen of Sauripterus taylori (Sarcopterygii, Osteichthyes) from the Famennian Catskill Formation of North America. J Vert Paleontol 24:26–40

99. Coates MI (1996) The Devonian tetrapod Acanthostega gunnari Jarvik: postcranial anatomy, basal interrelationships and patterns of skeletal evolution. Trans Roy Soc Edin Ear Sci 87:363–421

100. Kherdjemil Y, Kmita M (2018) Insights on the role of hox genes in the emergence of the pentadactyl ground state. Genesis 56:e23046

101. Leite-Castro J, Beviano V, Rodrigues P, Freitas R (2016) HoxA Genes and the Fin-to-Limb Transition in Vertebrates. J Dev Biol 4:10. https://doi.org/10.3390/jdb4010010

102. Hara Y, Yamaguchi K, Onimaru K, et al (2018) Shark genomes provide insights into elasmobranch evolution and the origin of vertebrates. Nat Ecol Evol 2:1761–1771. https://doi.org/10.1038/s41559-018-0673-5

103. Salazar-Ciudad I, Newman SA, Solé R V. (2001) Phenotypic and dynamical transitions in model genetic networks I. Emergence of patterns and genotype-phenotype relationships. Evol Dev 3:84–94. https://doi.org/10.1046/j.1525-142X.2001.003002084.x

104. Verd B, Clark E, Wotton KR, et al (2018) A damped oscillator imposes temporal order on posterior gap gene expression in Drosophila. PLoS Biol 16:e2003174. https://doi.org/10.1371/journal.pbio.2003174

105. Jiménez A, Cotterell J, Munteanu A, Sharpe J (2015) Dynamics of gene circuits shapes evolvability. Proc Natl Acad Sci 112:2103–2108. https://doi.org/10.1073/pnas.1411065112

106. Onimaru K, Nishimura O, Kuraku S (2018) A regulatory-sequence classifier with a neural network for genomic information processing. bioRxiv 355974. https://doi.org/10.1101/355974

107. Quang D, Xie X (2016) DanQ: A hybrid convolutional and recurrent deep neural network for quantifying the function of DNA sequences. Nucleic Acids Res 44:e107. https://doi.org/10.1093/nar/gkw226

108. Zhou J, Troyanskaya OG (2015) Predicting effects of noncoding variants with deep learning–based sequence model. Nat Methods 12:931–934. https://doi.org/10.1038/nmeth.3547